\documentclass[11pt]{article}
\usepackage{amsthm}
\usepackage{xspace}
\usepackage{fullpage}
\usepackage[utf8]{inputenc}
\usepackage{graphicx}
\usepackage{amsmath,amssymb}
\usepackage{float}
\usepackage{amsfonts}
\usepackage[ruled,noend,linesnumbered]{algorithm2e}
\usepackage{enumerate,comment}
\usepackage{url}
\usepackage[dvipsnames,usenames]{color}
\usepackage[colorlinks=true,urlcolor=Blue,citecolor=Green,linkcolor=BrickRed]{hyperref}
\usepackage[usenames,dvipsnames]{xcolor}
\usepackage{thm-restate}
\usepackage[capitalise]{cleveref}
\usepackage{todonotes}
\usepackage[noadjust]{cite}

\newcommand{\ceil}[1]{\lceil #1 \rceil}
\newcommand{\floor}[1]{\lfloor #1 \rfloor}

\newcommand{\sub}{\subseteq}
\newcommand{\sm}{\setminus}
\newcommand{\tlce}{T_{\mathrm{LCE}}}
\newcommand{\str}{\mathrm{str}}
\newcommand{\lca}{\mathrm{lca}}
\newcommand{\LCE}{\mathrm{LCE}}

\newcommand{\pred}{\mathrm{pred}}

\newcommand{\Oh}{\mathcal{O}}

   \newtheorem{theorem}{Theorem}[section]
   \newtheorem{corollary}[theorem]{Corollary}
   \newtheorem{lemma}[theorem]{Lemma}
   \newtheorem{observation}[theorem]{Observation}
   \newtheorem{fact}[theorem]{Fact}
   \theoremstyle{definition}   
   
   \usepackage{authblk}
   \theoremstyle{remark}
   
   \newtheorem*{claim}{Claim}

\begin{document}

\title{Sparse Suffix Tree Construction in Optimal Time and Space}

\author[1]{Paweł Gawrychowski\thanks{Work done while the author held a post-doctoral position at Warsaw Center of Mathematics and Computer Science.}}
\author[2]{Tomasz Kociumaka\thanks{Supported by Polish budget funds for science in 2013-2017 as a research project under the `Diamond Grant' program.}}

\affil[1]{University of Haifa, Israel}
\affil[2]{Institute of Informatics,
    University of Warsaw, Poland}
\affil[ ]{\texttt{\{gawry,kociumaka\}@mimuw.edu.pl}}

\date{}
\maketitle
\thispagestyle{empty}
\setcounter{page}{0}

\begin{abstract}
Suffix tree (and the closely related suffix array) are fundamental structures capturing all substrings
of a given text  essentially by storing all its suffixes in the lexicographical order.
In some applications,
such as sparse text indexing, we work with a subset of $b$ interesting suffixes, which are stored
in the so-called sparse suffix tree. Because the size of this structure is $\Theta(b)$,
it is natural to seek a construction algorithm using only $O(b)$ words of space assuming
read-only random access to the text.
We design a linear-time Monte Carlo algorithm for this problem,
hence resolving an open question explicitly stated by Bille et al. [TALG 2016].
The best previously known algorithm by I et al. [STACS 2014] works in $\Oh(n\log b)$ time.
As opposed to previous solutions, which were based
on the divide-and-conquer paradigm, our solution proceeds in $n/b$ rounds. In the $r$-th round,
we consider all suffixes starting at positions congruent to $r$ modulo $n/b$. By
maintaining rolling hashes, we can lexicographically sort all interesting suffixes
starting at such positions, and then we can merge them with the already considered suffixes.
For efficient merging, we also need to answer LCE queries efficiently (and in small space). By plugging
in the structure of Bille et al. [CPM 2015]  we obtain $\Oh(n+b\log b)$ time complexity.
We improve this structure by a recursive application of the so-called difference covers,
which then implies a linear-time sparse suffix tree construction algorithm.

We complement our Monte Carlo algorithm with a deterministic verification procedure.
The verification takes $\Oh(n\sqrt{\log b})$ time, which improves upon
the bound of $\Oh(n\log b)$ obtained by I et al. [STACS 2014]. This is obtained
by first observing that the pruning done inside the previous solution
has a rather clean description using the notion of graph spanners with small multiplicative
stretch. Then, we are able to decrease the verification time by applying
difference covers twice. Combined with the Monte Carlo algorithm, this gives us an $\Oh(n\sqrt{\log b})$-time and $\Oh(b)$-space Las Vegas algorithm. 
\end{abstract}

\newpage

\section{Introduction}\label{sec:introduction}
In many if not all algorithms operating on texts one needs a compact representation of all substrings.
A well-known data structure capturing all substrings of a given text is the suffix tree, which is a compacted trie storing all suffixes. 
The size of the suffix tree is linear in the length of the text and it provides efficient indexing,
that is, locating all occurrences of a given pattern.
The first linear-time suffix tree construction algorithm was given by Wiener~\cite{DBLP:conf/focs/Weiner73}. 
Later, McCreight~\cite{DBLP:journals/jacm/McCreight76} provided a simpler procedure, 
and Ukkonen showed a different approach that allows the text to be maintained under appending characters~\cite{DBLP:journals/algorithmica/Ukkonen95}. 
All these algorithms work in linear time assuming constant-size alphabet. 
However, such assumption is not always justified.
Farach developed a different construction method based on the divide-and-conquer paradigm
that takes only linear time as long as the alphabet is linear-time sortable~\cite{DBLP:conf/focs/Farach97}.
In particular, his algorithm works in linear time for polynomially-bounded integer alphabets.

While the suffix tree provides a lot of information about the structure of the text
and hence is a very convenient building block in more complicated algorithms, its large memory
footprint is often prohibitive in practice. Hence there has been a lot of interest in suffix arrays~\cite{DBLP:journals/siamcomp/ManberM93}.
A suffix array is just a lexicographically sorted list of all suffixes of the text. The list is usually
augmented with the so-called LCP table, which stores the length of the longest common prefix
of every two adjacent suffixes. This leads to a very memory efficient representation that is still
capable of providing enough information about the text to replace suffix trees in all applications
with no or very small penalty in the time complexity~\cite{DBLP:journals/jda/AbouelhodaKO04}.
Furthermore, a suffix array can be constructed in linear time for any linear-time sortable
alphabet with a simple and practical algorithm~\cite{DBLP:journals/jacm/KarkkainenSB06}.

Even though the suffix array occupies linear space when measured in words, this might be
larger than the encoding of the text. This started a long line of work on compressed suffix
arrays~\cite{DBLP:conf/focs/FerraginaM00,DBLP:conf/soda/GrossiGV03,DBLP:journals/siamcomp/GrossiV05}, which take space proportional to the entropy
of the text. However, in some applications even smaller space usage is desired.
In particular, in the last few years there has been a lot of interest among the string algorithms community
in designing sublinear space solutions, where one assumes a read-only random
access to the input text and measures the working space. Of course, the running time should
still be linear or close to linear. 

A natural idea for text indexing in sublinear space is to use some additional knowledge about the structure
of the queries to consider only a (small) subset of all suffixes of the text, and provide indexing
only for occurrences starting at the corresponding positions.
This was first explored by Kärkkäinen and
Ukkonen~\cite{DBLP:conf/cocoon/KarkkainenU96}, who introduced the sparse suffix tree.
They showed that the evenly space sparse suffix tree, which is the compacted trie storing
every $k$-th suffix of the text, can be constructed in linear time and working space
proportional to the number of suffixes. However, the question of construct a general
sparse suffix tree (or sparse suffix array, which is a lexicographically sorted list of the
chosen subset of suffixes together with their LCP information) for an arbitrary subset of
$b$ suffixes of a text $T[1..n]$ using $\Oh(b)$ working space, remained open.
Recently, Bille et al.~\cite{DBLP:conf/icalp/BilleFGKSV13} were able to make a significant progress
towards resolving this question by developing an $\Oh(n\log^2b)$-time Monte Carlo algorithm.
They also provided a verification
procedure implying an $\Oh(n\log^2b+b^2\log b)$ Las Vegas algorithm.
Then, I et al.~\cite{DBLP:conf/stacs/IKK14} improved the complexity to $\Oh(n\log b)$ (both for Monte Carlo and Las Vegas randomization).
They also gave an $\Oh(n)$-time Monte Carlo solution using $\Oh(b\log b)$ space.
Very recently, Fischer et al.~\cite{DBLP:conf/latin/0001IK16} gave an $\Oh(n\sqrt{\log n}+b\log b\log n \log^*n)$-time 
and $\Oh(b)$-space deterministic algorithm in a stronger model of rewritable text which needs to be restored before
termination.

Another natural problem in the model of read-only random access to the text is LCE queries,
where we are to preprocess a text subject to queries $\LCE(i,j)$ returning the longest common prefix
of two suffixes $T[i..]$ and $T[j..]$.
Several trade-off between query time, data structure size, construction time and space usage have been obtained~\cite{DBLP:journals/jda/BilleGSV14,DBLP:conf/cpm/BilleGKLV15,DBLP:conf/cpm/TanimuraIBIPT16}.
The queries are typically deterministic, but construction algorithms range from Monte Carlo randomization via Las Vegas randomization to deterministic solutions.
State-of-the-art Monte Carlo data structures of $\Oh(b)$ size 
have $\Oh(n)$-time and $\Oh(b)$-space construction with $\Oh(n/b+\log b)$-time queries,
or $\Oh(n\log b)$-time and $\Oh(b)$-space construction with $\Oh(n/b)$-time queries~\cite{DBLP:conf/cpm/BilleGKLV15}.

\paragraph{Our contribution.}
We design an $\Oh(n)$-time Monte Carlo algorithm for sorting an arbitrary subset of $b$ suffixes
of a text $T[1..n]$ using $\Oh(b)$ working space. 
We also show how to verify the answer in $\Oh(n\sqrt{\log b})$ time and $\Oh(b)$ working space,
which implies a Las Vegas algorithm with such complexity.
Hence, for Monte Carlo algorithms we close the problem, while for Las Vegas algorithm we are able to make a substantial
progress towards the desired linear time complexity.

As an auxiliary result, we also develop an $\Oh(b)$-space LCE data structure
with $\Oh(n/b)$-time queries and $\Oh(n)$-time and $\Oh(b)$-space Monte Carlo construction. 

\paragraph{Model.} We are given read-only random access to a text $T[1..n]$ consisting of characters from $\Sigma=\{1,2,\ldots,n^{\Oh(1)}\}$.
 We assume the standard word RAM model with word size $\Theta(\log n)$, where basic arithmetic and bit-wise operations on $\Oh(\log n)$-bit integers take constant time.
 Our randomized algorithms succeed with high probability, i.e., $1-n^{-c}$ for any user-specified constant $c$. 

\paragraph{Previous and our techniques.}
The Monte Carlo algorithm of I et al.~\cite{DBLP:conf/stacs/IKK14} is based on the notion of $\ell$-strict sparse suffix trees. 
Intuitively, they are approximate variants of the sparse suffix tree operating on blocks of length $\ell$ rather with single-character precision.
The algorithm starts with a trivial $n$-strict sparse suffix tree and performs $\ceil{\log n}$ steps, each of which halves the block length.

Our algorithm, described in \cref{sec:montecarlo}, performs just two steps.
Its intermediate result, the \emph{coarse compacted trie}, is basically the same as the $\ceil{n/b}$-strict sparse suffix tree. 
The second phase, building the sparse suffix tree from the coarse compacted trie, is relatively easy.
For the more challenging first step, we employ Karp--Rabin fingerprints stored in a rolling fashion.
More precisely, we proceed in $\ceil{n/b}$ rounds; in the $r$-th round, we store fingerprints of all suffixes starting at positions congruent to $r$ modulo $\ceil{n/b}$.
This way, while inserting such a suffix to the coarse compacted trie, we can guarantee that almost all fingerprints needed can be accessed in constant time.
The last step of every insertion reduces to a longest common extension (LCE) query.
If we apply a recent data structure by Bille et al.~\cite{DBLP:conf/cpm/BilleGKLV15}, the total running time becomes $\Oh(n+b\log b)$. 

To obtain $\Oh(n)$ time, we improve the LCE query time to $\Oh(n/b)$.
Bille et al.~\cite{DBLP:conf/cpm/BilleGKLV15} already applied difference covers for that purpose, 
but this was at the expense of superlinear preprocessing time since a certain sparse suffix tree had to be constructed.
Our main insight is that one can build a sequence of larger and larger sparse suffix trees,
each of them providing faster LCE queries used to construct the next sparse suffix tree.

Our complementary result, an $\Oh(n\sqrt{\log b})$-time Las Vegas algorithm,
is provided in \cref{sec:lasvegas}.
In short, our solution is based on an $\Oh(n\log^2 b)$-time algorithm by I et al.~\cite{DBLP:conf/stacs/IKK14} for verifying $b$ substring equations.
We provide a few improvements by exploiting a slightly cleaner (though more complex) formalization of the underlying ideas.
First, we avoid eagerly checking some constraints, which lets us reduce the running time to $\Oh(n\log b + b\log^2 n)$.
Next, we use difference covers to restrict the set of starting positions of fragments involved in equations;
this technique independently speeds up two steps of the algorithm.

\section{Preliminaries}
We consider finite strings over an integer alphabet $\Sigma=\{1,\ldots,n^{\Oh(1)} \}$.
For a string $T=T[1]\cdots T[n]$, its \emph{length} is $|T|=n$.
For $1\le i \le j \le n$, a string $T[i] \cdots T[j]$ is called a \emph{substring} of $T$.
By $T[i..j]$ we denote its occurrence at position $i$, called a \emph{fragment} of~$T$.
A fragment with $i=1$ is called a \emph{prefix} (also denoted $T[..j]$)
and a fragment with $j=n$ is called a \emph{suffix} (denoted $T[i..]$).

\renewcommand{\S}{\mathcal{S}}

\paragraph{Tries and compacted tries.}
Recall that a \emph{trie} is a rooted tree whose nodes correspond to prefixes of strings in a given set of strings $\S$.
The prefix corresponding to a node $u$ is denoted $\str(u)$, and the node $u$ is called the \emph{locus} of $\str(u)$.
We extend this notion as follows:
the locus of an arbitrary string $s$ is the node $u$ such that $\str(u)$ is prefix of $s$ and
and $|\str(u)|$ is maximized.

The parent-child relation in the trie is defined so that the root is the locus of $\varepsilon$,
while the parent~$u$ of a node $v$ is the locus of $\str(v)$ without the last character.
This character is the \emph{label} of the edge from $u$ to $v$.
The order on the alphabet naturally yields an order on the edges outgoing from any node of the trie,
so tries are often assumed to be ordered rooted trees.

A node $u$ is \emph{branching} if it has at least two children and \emph{terminal} if $\str(u)\in \S$.
A \emph{compacted trie} is obtained from the underlying trie by dissolving all nodes except
the root, branching nodes, and terminal nodes. The dissolved nodes are called \emph{implicit} while the preserved nodes are called \emph{explicit}.
The compacted trie takes $\Oh(|\S|)$ space provided that edge labels are stored as pointers to fragments of strings in $\S$.
In some applications, the first character is kept explicitly, however.

Note that the suffix tree of a string $T$ is precisely the compacted trie of the set of all suffixes of $T$.
Similarly, the sparse suffix tree of an arbitrary set $B$ of suffixes of $T$ is the compacted trie of $B$.
Given the lexicographic order on $\S$ along with the lengths of the longest common prefixes between any two consecutive (in this order) elements of $\S$,
one can easily compute the compacted trie in $\Oh(|\S|)$ time; see e.g.~\cite{AlgorithmsOnStrings}.
Thus, the problem of constructing the sparse suffix tree is equivalent to that of building the sparse suffix array along with the LCP values.

\paragraph{LCA queries.}
For two nodes $u,v$ of a trie, we denote their lowest common ancestor by $\lca(u,v)$.
Since $\str(\lca(u,v))$ is the longest common prefix of $\str(u)$ and $\str(v)$,
the data structures for LCA queries are often applied to efficiently determine longest common prefixes.

\begin{lemma}[\cite{DBLP:journals/siamcomp/HarelT84,LCA}]\label{lem:lca}
The compacted trie of a set of strings $\S$ can be preprocessed in $\Oh(|\S|)$ time to
compute the length of the longest common prefix of any two strings in $\S$ in constant time.
\end{lemma}
\paragraph{Karp--Rabin fingerprints.}
For a prime number $p$, an integer $x\in \mathbb{Z}_p$, 
the Karp--Rabin fingerprint~\cite{DBLP:journals/ibmrd/KarpR87} of a string $w$ is
$(\sum_{i=1}^n w[i]\cdot x^{i-1})\bmod{p}$.
For efficiency, we augment it as follows:
$$\phi(w)=\left(\left(\sum_{i=1}^n w[i]\cdot x^{i-1}\right) \bmod p,\,  x^{|w|} \bmod p,\, x^{-|w|} \bmod p,\,|w|\right).$$

\begin{observation}\label{obs:fingerprint_cmp}
Let $u,v,w$ be strings such that $uv = w$.
Given two out of three fingerprints $\phi(u), \phi(v),\phi(w)$, the third one can be computed in constant time.
\end{observation}

For a text $T$, we say that the fingerprints are collision-free
if $\phi(T[i..j])=\phi(T[i'..j'])$ implies $T[i..j]=T[i'..j']$.
Randomization lets us construct such fingerprints with high probability:
\begin{fact}\label{fct:collision_free}
Let $T$ be a text of length $n$ over alphabet $\Sigma$, and let $p$ be a prime number such that $p\ge \max(|\Sigma|, n^{3+c})$.
If $x$ is uniformly random, $\phi$ is collision-free with probability at least $1-n^{-c}$.
\end{fact}

This is the only source of randomization in this paper. The original argument
by Karp and Rabin~\cite{DBLP:journals/ibmrd/KarpR87} used random $p$ and fixed $x$,
but we use the more modern approach; see e.g.~\cite{DBLP:conf/icalp/BilleFGKSV13,DBLP:conf/cpm/BilleGKLV15,DBLP:conf/stacs/IKK14}.

\paragraph{LCE queries.} 
For a text $T$ of length $n$ and two positions $i,j$ ($1\le i,j\le n$),
we define $\LCE(i,j)$ as the length of the longest common prefix of $T[i..]$ and $T[j..]$. 
We use the following recent result as a building block in our algorithms:
 
\begin{lemma}[Bille et al.~\cite{DBLP:conf/cpm/BilleGKLV15}]\label{lem:lce}
Given read-only random access to a text $T$ of length $n$ and a parameter
$b$, $1\le b\le n$, it is possible to construct in $\Oh(n)$ time and $\Oh(b)$ space
a structure of size $\Oh(b)$,
which answers LCE queries in $\Oh(n/b+\log(b\cdot \ell/n))=\Oh(n/b+\log b)$ time,
where $\ell$ is the result of the query.
The data structure assumes collision-free Karp--Rabin fingerprints.
\end{lemma}

\paragraph{Difference covers.}
We say that a set $DC\sub \{1,\ldots,n\}$ is a \emph{$t$-difference-cover} (of $\{1,\ldots,n\}$)
if for each $i,j\in \{1,\ldots,n-t\}$ there is a value $h(i,j)$, $0\le h(i,j)<t$,
such that $i+h(i,j)\in DC$ and $j+h(i,j)\in DC$. We say that $DC$ can be \emph{indexed}
efficiently, if there is a bijection $f : DC \to \{1,\ldots,|DC|\}$ such that $f$ and $f^{-1}$ are computable in constant time.

\begin{lemma}[Maekawa~\cite{Maekawa}, Burkhard and Kärkkäinen~\cite{BurkhardtEtAl2003}]\label{lem:cover}
For every positive integers $t$ and $n$, $t\le n$, there exists a $t$-difference-cover $DC$ of size $\Oh(\frac{n}{\sqrt{t}})$,
such that $h$ can be evaluated in constant time and $DC$ can be efficiently indexed.
\end{lemma}

\section{Monte Carlo Algorithm}\label{sec:montecarlo}
In this section, we provide an $\Oh(n)$-time and $\Oh(b)$-space Monte Carlo algorithm
for computing the sparse suffix tree. 
In \cref{sec:coarse}, we introduce the coarse compacted trie, which is an intermediate
byproduct of our algorithm,
and we show how to use it to build the sought sparse suffix tree.
Then we concentrate on computing the coarse compacted trie. 
\cref{sec:constr} provides an $\Oh(n+b\log b)$-time algorithm whose bottleneck are LCE queries.
We overcome this in \cref{sec:lce} by providing an improved data structure for LCE queries.
We conclude the exposition in \cref{sec:sum}.

\subsection{Coarse Compacted Tries}\label{sec:coarse}
The \emph{coarse compacted trie} of a collection of strings is defined as follows. 
We conceptually partition each string into blocks consisting of $\ceil{n/b}$ characters (the last block might be shorter).
Then we consider each block to be a single supercharacter, and we form the compacted trie of the resulting set of strings. 
We use fingerprints to represent the supercharacters; hence, the order on the edges outgoing from the same node
in a coarse compacted trie corresponds to the order of the fingerprints of the first block on every edge, not to the lexicographical order of the blocks. 

Below, we show that the sought sparse suffix tree can be quite easily derived from the corresponding coarse compacted trie.
The key building block is sorting $\Oh(b)$ strings of length $\ceil{n/b}$.
For $b\le \sqrt{n}$, a simple comparison-based algorithm achieves $\Oh(n)$ time and $\Oh(b)$ space complexity. 

\begin{fact}\label{fct:sort_small}
Given random access to $b$ strings of length $\ell$, the strings can be sorted in $\Oh(b(b+\ell))$ time and $\Oh(b)$ space.
\end{fact}
\begin{proof}
We shall prove that a single string $S$ can be inserted into a sorted array of $b$ strings in $\Oh(b+\ell)$ time. 
We scan the consecutive letters of $S$. 
Having read $S[j]$, we maintain a partition of the strings $T$ in the array into three classes depending on whether $T[..j]$ is smaller, equal, or greater than $S[..j]$.
Note that these classes form consecutive ranges. 
Moreover, in order to update the partition, it suffices to scan the `equal' class  from both ends and remove all leading strings $T$ satisfying $T[j+1]<S[j+1]$,
and all trailing strings $T$ satisfying $T[j+1]>S[j+1]$. 
The running time is proportional to the number of strings removed plus $\Oh(1)$,
which gives $\Oh(b+\ell)$ in total over all steps.
After we scan the whole $S$, the partition determines the position where it should be inserted.
\end{proof}

For $b\ge \sqrt{n}$, we have enough space to use an algorithm based on counting sort.

\begin{fact}\label{fct:sort}
Assume we are given random access to $b$ strings of length $\ell$ over alphabet $\{1,\ldots,\sigma\}$.
For any positive integer $S \le \sigma$, the strings can be sorted in $\Oh((b+S)\ell \log_{S}\sigma)$ time and $\Oh(b+S)$ space.
\end{fact}
\begin{proof}
We treat every character as a $\log \sigma$-bit integer
and partition it into chunks of $\floor{\log S}$ bits.
We radix sort the resulting collection of $b$ strings of length $\Theta(\ell \log_{S}\sigma)$
over a smaller alphabet $\{1,2,\ldots,S\}$.
With $\Theta(\ell \log_{S}\sigma)$ iterations of counting sort, we obtain the claimed bounds.
\end{proof} 

Now, we can give the procedure building the compacted trie from of the coarse compacted trie.

\begin{lemma}
\label{lem:uncoarse}
Given the coarse compacted trie of a set of $b$ suffixes, we can construct
their compacted trie in $\Oh(n)$ time and $\Oh(b)$ space.
\end{lemma}

\begin{proof}
Consider a branching node $u$ of the coarse compacted trie and let $v_1,v_2,\ldots,v_k$
be its implicit children, that is, $(u,v_i)$ is an edge labelled with a single supercharacter representing a string $s_i$ of length $\ceil{n/b}$.
To obtain the compacted trie, we remove the edges $(u,v_i)$ and paste the compacted trie
of the strings $s_1,s_2,\ldots,s_k$ to connect $u$ with $v_1,v_2,\ldots,v_k$.
Such a trie is easy to construct in $\Oh(k\ceil{n/b})$ time by inserting strings in the lexicographic order: 
we follow the rightmost path while its label matches the inserted string, 
and we create a new branch as soon as it does not match. 
Hence, we only need to show how sort the strings $s_1,s_2,\ldots,s_k$.

We gather such collections of strings from all branching nodes, sort the disjoint union of all these collections, and then recover the sorted collections with a single scan. 
Note that the total number of strings to be sorted is bounded by the number of edges in the coarse compacted trie, which is $\Oh(b)$.
If $b\le \sqrt{n}$, we sort using \cref{fct:sort_small}, while for $b>\sqrt{n}$ we apply \cref{fct:sort} with $\sigma=n^{\Oh(1)}$ and $S=b$.
In both cases, sorting is performed is $\Oh(n)$ time and $\Oh(b)$ space.
This is also the overall time and space necessary to construct compacted tries based on the sorted collections.
\end{proof}

\subsection{Construction of Coarse Compacted Tries}\label{sec:constr}
Again, we provide slightly different construction algorithms for $b\leq \sqrt{n}$ and $b> \sqrt{n}$.
Both algorithms use \cref{lem:lce} and rolling Karp--Rabin fingerprints, which we describe below.
Moreover, while constructing the coarse compacted trie, at each explicit node $v$ we always store the fingerprint of the corresponding string $\str(v)$. 
Similarly, each edge stores the fingerprint of the first block of its label, and a reference to a fragment of $T$ representing the whole label.
For each node, the outgoing edges are kept in a doubly-linked list (ordered by the fingerprints of their first blocks).

For an integer $r$, we say a position $i$ is \emph{$r$-aligned} if $i \equiv r \pmod{\ceil{n/b}}$.
A fragment $T[i..j]$ is called an \emph{$r$-aligned fragment} of $T$ if $i$ is an $r$-aligned position
and $j=n$ or $j+1$ is also $r$-aligned. 
The key idea of our algorithm is to proceed in $\ceil{n/b}$ rounds
so that in the $r$-th round, while inserting $r$-aligned suffixes,
we can compute the fingerprint of any $r$-aligned fragment in constant time.
More formally, we denote the set of all $r$-aligned suffixes by $S_r$, and we define a component $\Phi_r$ consisting of the fingerprint of every suffix in $S_r$, 
as well as the fingerprint of the whole text $T$.
The following fact, stating all the necessary properties of $\Phi_r$, easily follows from \cref{obs:fingerprint_cmp}.
\begin{fact}\label{fct:rolling}
The component $\Phi_1$ can be constructed in $\Oh(n)$ time and $\Oh(b)$ space.
Moreover, given $\Phi_r$, the  fingerprint of any $r$-aligned fragment can be computed in constant time,
and $\Phi_{r+1}$ can be constructed in $\Oh(b)$ time and space.
\end{fact}

\subsubsection{Small $b$}
\label{sec:smallb}
First, we show how to construct coarse compacted trie for a set  of $b\le \sqrt{n}$ suffixes.
Our procedure actually works for an arbitrary number of suffixes, but it is too slow for $\omega(\sqrt{n})$ suffixes.

\begin{theorem}
\label{thm:smallb}
The coarse compacted trie of set of $b$ suffixes of a text $T$ of length $n$ can be computed in $\Oh(n+b^2)$ time.
\end{theorem}
\begin{proof}
We proceed in rounds corresponding to $r=1,\ldots,\ceil{n/b}$. 
In the $r$-th round, we insert $r$-aligned suffixes $T[i..]$ one by one.
For this, we use $\Phi_r$ and the LCE-structure from \cref{lem:lce}.

We locate the locus of $T[i..]$ in the current trie by a traversal starting at the root.
If we are currently at an explicit node $u$, we use $\Phi_r$ to obtain the fingerprint of the next supercharacter to be followed. 
Next, we scan the edges going out of $u$ comparing that fingerprint with the ones store with the edges.
If none of them matches, then $u$ is the locus and we insert a new leaf with $u$ as its parent.
Otherwise, we have selected an edge leading to a child $v$ of $u$.
We check if the locus is in the subtree of $v$ by comparing $\phi(\str(v))$ with the fingerprint of the corresponding ($r$-aligned) prefix of $T[i..]$.
If so, we continue the traversal at $v$.
Otherwise, we know that the locus is an implicit node on the edge  $(u,v)$.
In this case, we use one LCE query (rounded down to a multiple of $\ceil{n/b}$) to calculate the exact position of the locus on the edge, and we attach a new leaf there. 
Then, we also need to spend $\Oh(n/b)$ time to compute the fingerprint of the edge created by subdividing $(u,v)$ at the locus.
All other fingerprints stored in the new nodes and edges are computed in constant time since the corresponding fragments are $r$-aligned.

The trie is of size $\Oh(b)$, so in a single traversal we visit at most $\Oh(b)$ explicit nodes and scan $\Oh(b)$ edges in total,
spending $\Oh(1)$ time at each of them.
The last step requires $\Oh(n/b+\log b)$ time for an LCE query and fingerprint computation,
so the overall insertion time is $\Oh(b+n/b+\log b)=\Oh(b+n/b)$.
Summing over all the rounds, this is $\Oh(n+b^2)$ (including the time to maintain $\Phi_r$ and
build the LCE-structure). Space consumption remains $\Oh(b)$ throughout the algorithm.
\end{proof}

\subsubsection{Large $b$}\label{sec:mediumb}
For larger $b$, instead of processing suffixes one by one, we insert all suffixes from $B_r = B\cap S_r$ in bulk. 
Our insertion algorithm requires the coarse compacted trie of $B_r$,
which is obtained from the coarse compacted trie of $S_r$.

\begin{lemma}\label{lem:sr}
For $b=n^{\Omega(1)}$, given $\Phi_r$ the coarse compacted trie of $S_r$ can be computed in $\Oh(b)$ time and space.
\end{lemma} 
\begin{proof}
For each $r$-aligned position $i$, we define its \emph{block} as $T[i..\min(n,i+\ceil{n/b}-1)]$.
Then, we define a string $T_r$ where $T_r[j]$ is the fingerprint of the block corresponding to the $j$-th leftmost $r$-aligned position in $T$.
In other words, $T_r$ is obtained from $T$ by partitioning $T[r..]$ into blocks of size $\ceil{n/b}$ and replacing each block by its fingerprint.
Observe that this partition coincides with the partitions of $r$-aligned suffixes in the definition of the coarse compacted trie of $S_r$.
Thus, the coarse compacted trie of $S_r$ is precisely the suffix tree of $T_r$.

The first step of our construction algorithm is to compute $T_r$. 
Using $\Phi_r$, this takes $\Oh(b)$ time since blocks are $r$-aligned fragments.
Then, we sort the letters of $T_r$ using \cref{fct:sort} with $\ell=1$, $\sigma=n^{\Oh(1)}$, and $S=b$.
This way, using $\Oh(b\log_b n)=\Oh(b)$ time and $\Oh(b)$ space, each character of $T_r$ can be replaced by its rank, which is at most $b$.
After such normalization, we construct the suffix tree of $T_r$ in $\Oh(b)$ time and space. 
Finally, we replace the normalized characters on the edges by the original $\Oh(\log n)$-bit fingerprints.
\end{proof}

Next, we implement the bulk insertion procedure. 
Note that coarse compacted trie of $B_r$ can be extracted in $\Oh(b)$ time from the coarse compacted trie of $S_r$.
We define $B_{<r} = \bigcup_{j=1}^{r-1} B_j$.

\begin{lemma}
\label{lem:merge}
Assume that we have access to $\Phi_r$ and an LCE-structure with $\tlce$ query time.
Given the coarse compacted trie for $B_{<r}$ and the coarse compacted trie for $B_r$, 
we can construct the coarse compacted trie for $B_{<r}\cup B_r$
in $\Oh(b+|B_r|(\tlce+n/b))$ time and $\Oh(b)$ space.
\end{lemma}

\begin{proof}
Intuitively, our aim is to traverse the Euler tour of the resulting coarse compacted trie of $B_{<r}\cup B_r$.
More precisely, we process consecutive strings $s_1,\ldots,s_k \in B_r$ and for each $s_i$ we find the locus
of $s_i$ in the trie of $B_{<r}\cup \{s_1,\ldots,s_{i-1}\}$ and then add $s_i$ to the trie.
Strings $s_i$ are processed according to the coarse lexicographic order, so we only need to move forward on the Euler tour
to reach the next locus; actually, we can start from the new terminal node (representing $s_{i-1}$), possibly skipping some part of the Euler tour.

While visiting an edge $(u,v)$ whose label starts with a supercharacter $c$, we shall find out if the locus of $s_i$ is an implicit node on that edge.
An equivalent condition is that $\str(u)$ is a prefix of $s_i$ followed by a block represented by $c$, but $\str(v)$ is not a prefix of $s_i$. 
For this, we compute the fingerprints of the appropriate fragments of $s_i$. Since $s_i\in S_r$, these fragments are $r$-aligned in $T$,
so the fingerprints are determined in $\Oh(1)$ time.
If the locus turns out to be on the current edge, we make an LCE query to determine its exact depth; 
the result needs to be rounded down to full blocks. 
Next, we introduce a new explicit node $v'$ on the edge $(u,v)$ and a terminal node representing $s_i$.
Fingerprints stored at the new nodes and edges can be computed in $\Oh(1)$ time, except for the one at $(v',v)$, which takes $\Oh(n/b)$ time
as the corresponding block may not be $r$-aligned.

Similarly, while at an explicit node $v$, we first check if $\str(v)$ is a prefix of $s_i$;
otherwise, the subtree of $v$ can be ignored. 
Next, we compute the supercharacter $c$ representing the block of $s_i$
following $\str(v)$ and compare it to the labels of edges from $v$ to its children.
Note that a vertex with $d$ children appears on the Euler tour $d+1$ times, so during any visit to $v$ we need to consider at most two edges
whose labels are the lower bound and the upper bound on $c$. 
If $c$ turns out to be within the range, we simply introduce the leaf representing $s_i$ as a child of $v$.
The new edge (labeled with $c$) is inserted between the two considered.

Hence, in either case  it takes $\Oh(1)$ time to make progress traversing the Euler tour, and at most $\Oh(\tlce+n/b)$
time to extend the trie. This sums up to $\Oh(b+|B_r|(\tlce+n/b))$ in total.
\end{proof}

\begin{theorem}
\label{thm:mediumb}
Assume that we have access to LCE-structure with $\tlce$ query time.
The coarse compacted trie of any set $B$ of $b=n^{\Omega(1)}$ suffixes can be
computed using $\Oh(n+b\cdot \tlce)$ time and $\Oh(b)$ space.
\end{theorem}

\begin{proof}
We iterate over $r=1,\ldots,\ceil{n/b}$ while maintaining $\Phi_r$. By \cref{fct:rolling}, this takes $\Oh(n)$ time and $\Oh(b)$ space in total.
In the $r$-th round, we apply \cref{lem:sr} to compute the coarse compacted trie of $B_r$,
and then we use \cref{lem:merge} to merge it with the coarse compacted trie of $B_{<r}$ to obtain the coarse compacted
trie of $B_{<\,r+1}$.
The total cost of all applications of \cref{lem:merge} is
\[ \sum_{r=1}^{\ceil{n/b}}  \Oh(b+|B_r|(n/b+\tlce)) = \Oh(n + b\cdot(n/b+\tlce))=\Oh(n+b\cdot \tlce).\qedhere\]
\end{proof}

\begin{corollary}
\label{cor:slowMC}
For any set of $b$ suffixes of $T[1..n]$, the sparse suffix tree can be
computed using $\Oh(n+b\log b)$ time and $\Oh(b)$ space.
The resulting tree might be incorrect with probability $n^{-c}$ for a user-defined constant $c$.
\end{corollary}
\begin{proof}
The algorithm builds the coarse compacted trie of the suffixes.
If $b \leq \sqrt{n}$ is small, we use \cref{thm:smallb}.
Otherwise, we apply \cref{thm:mediumb} with \cref{lem:lce} to answer LCE queries.
This results in $\tlce =\Oh(n/b + \log b)$, so the total running time is $\Oh(n+b\log b)$.
Finally, we construct the sparse suffix tree using \cref{lem:uncoarse}. 
\end{proof}

\subsection{More efficient LCE queries}
\label{sec:lce}

In this section, we provide a faster data structure for LCE queries.
Our solutions takes $\Oh(b)$ space, has $\Oh(n)$ construction time and $\Oh(n/b)$ query time.
When used in \cref{thm:mediumb}, the running time immediately improves to $\Oh(n)$.

The main idea is similar to that by Bille et al. \cite{DBLP:conf/cpm/BilleGKLV15}: 
we use a sparse suffix tree for a $t$-difference cover to process LCE queries with answer at least $t$ in constant time.
Bille et al. used the algorithm by I et al.~\cite{DBLP:conf/stacs/IKK14} to build the tree, which resulted in $\Oh(n\log b)$ construction time.
We devise a recursive approach: we construct sparse suffix arrays for larger and larger difference covers,
using the previous one to speed up the construction of the next.
The largest of these difference covers consists of $\Oh(b^2/n)$ suffixes but we can still use $\Oh(b)$ space.
Thus, having relatively more space, we can apply a much simpler variant of the algorithm of \cref{thm:mediumb}.

Formally, at the $i$-th level of recursion, we have a set $B^{(i)}$ of $b_i = \Oh(b(\frac{b}{n})^i)$ suffixes,
which is a $t_i$-difference cover for $t_i=(\frac{n}{b})^{2(i+1)}$ obtained using \cref{lem:cover}.
The recursion terminates at $i=i_{\max}$ such that $(\frac{n}{b})^{2(i_{\max}+2)}>n$. Hence, $t_i\le n$ for $1\le i \le i_{\max}$.

\begin{lemma}[see~Section 3.6 in~\cite{DBLP:conf/cpm/BilleGKLV15}]\label{lem:dc}
After $\Oh(n)$-time and $\Oh(b)$-space preprocessing,
for every $i$, $1\le i \le i_{\max}$,
the sparse suffix tree of $B^{(i)}$ can be processed in $\Oh(b_i)$ time so that LCE queries can be answered in $\Oh(\frac{n}{b}+i\log \frac{n}{b})$ time.
\end{lemma}
\begin{proof}
We store the structure of \cref{lem:lce} and the component $\Phi_1$ (to test equality of fragments in $\Oh(n/b)$ time).
Given the sparse suffix tree of $B^{(i)}$, we build the data structure of \cref{lem:lca} for longest common prefix queries on $B^{(i)}$.
Next, we exploit the fact that  $B^{(i)}$ forms a difference cover.

To find $\LCE(p,q)$, we first test whether the result is at least $h(p,q)$.
This involves a single substring equality check.
If the answer is positive, we compute the result in constant time as $\LCE(p,q)=h(p,q)+\LCE(p+h(p,q),q+h(p,q))$
with the second summand determined using the component of \cref{lem:lca}.
Otherwise, we use the data structure of \cref{lem:lce}.
The running time is $\Oh(n/b + \log (b \cdot h(p,q) /n))=\Oh(n/b + \log((n/b)^{2i+1}))=\Oh(n/b + i\log(n/b))$.
\end{proof}
 
Next, we observe that the approach of \cref{lem:uncoarse} can also be used to process
a coarse compacted trie of $b'<b$ suffixes in the trie.
\begin{lemma}\label{lem:uncoarse3}
Let $b',b$ be positive integers such that $b'\le b\le n$ and $b'=n^{\Omega(1)}$.
Given the coarse compacted trie of an arbitrary set of $b'$ suffixes, we can
construct their compacted trie in $\Oh(nb'/b)$ time and $\Oh(b)$ space.
\end{lemma}

Recall that for $1 \le r \le \ceil{n/b}$, we defined $S_r$ as the set of $r$-aligned suffixes.
We denote $B^{(i)}_r = B^{(i)}\cap S_r$.  

\begin{lemma}\label{lem:generate}
The sparse suffix arrays of $B^{(i)}_r$ (for $1\le i \le i_{\max}$ and $1\le r \le \ceil{n/b}$) can be computed in $\Oh(n)$ time and $\Oh(b)$ space in total
if $b=n^{\Omega(1)}$.
\end{lemma}
\begin{proof}
First, we use \cref{lem:cover} to generate $B^{(i)}$ for each $i$. This takes $\Oh(\sum_{i} b_i)=\Oh(b^2/n)$ time and space.
Next, we lexicographically sort triples $(j \bmod \ceil{n/b},j,i)$ for $j\in B^{(i)}$ in $\Oh(b^2/n\log_b n+b)$ time and $\Oh(b)$ space
using \cref{fct:sort} for $\ell=3$, $S=b$, and $\sigma=n$.
Finally, we iterate over $r=1,\ldots \ceil{n/b}$ and construct the sparse suffix trees of $B^{(i)}_r$
based on the coarse compacted tries of $S_r$. Building the latter takes $\Oh(n)$ time and $\Oh(b)$ space across all iterations by \cref{lem:sr,fct:rolling}.
We also extend the tree with the data structure of \cref{lem:lca} for LCP queries.
We use the previously constructed list of triples to build an array indexed by elements $j\in S_r$
storing in each entry a list of integers $i$ such that $j\in B^{(i)}$.
Next, we extract coarse compacted tries of $B^{(i)}_r$ from the coarse compacted trie of $S_r$.
Finally, we use \cref{lem:uncoarse3} to obtain the sparse suffix arrays of $B^{(i)}_r$.
\end{proof}

\begin{theorem}\label{thm:lce}
There is a data structure of size $\Oh(b)$ which can be constructed in $\Oh(n)$
time and answers LCE queries in $\Oh(n/b)$ time. 
The data structure might be corrupted with probability $n^{-c}$ for a user-defined constant~$c$.

\end{theorem}
\begin{proof}
If $b \ge n/2$, we use the standard $\Oh(n)$-space data structure. 
Similarly, if $b \le n/\log n$, then the query time of the data structure of \cref{lem:lce}
is always $\Oh(n/b + \log b)=\Oh(n/b)$, so there is nothing to do as well.
Thus, we assume $n/\log n < b<n/2$.

First, we use \cref{lem:generate} to compute the sparse suffix arrays of $B^{(i)}_r$ for $1\le i \le i_{\max}$ and $1\le r \le \ceil{n/b}$.
Then, we iterate over $i=i_{\max},\ldots,1$ and, for each such $i$, merge the suffix arrays
of all $B^{(i)}_r$ to obtain the sparse suffix array of $B^{(i)}$.
This is easily achieved using $\Oh(b_i\log \frac{n}{b})$ LCE queries. 
We answer these queries using the component of \cref{lem:lce}  for $i=i_{\max}$
and \cref{lem:dc} (plugging in the sparse suffix tree constructed for $i+1$) for $i<i_{\max}$.
In either case, the cost of the LCE query is $\Oh(n/b + (i+1)\log (n/b))=\Oh(n/b+i\log (n/b))$.

The space consumption is clearly $\Oh(b)$ throughout the algorithm. 
The overall construction time is dominated by on LCE queries, leading to the total running time of
$$\Oh\left(\sum_{i=1}^{i_{\max}} b_i\left(\tfrac{n}{b}+i\log\tfrac{n}{b}\right)\right)=
\Oh\left(\sum_{i=1}^{i_{\max}} b_i \cdot \tfrac{n}{b}\cdot i \right)=
\Oh\left(n\sum_{i=1}^\infty i\cdot (\tfrac{b}{n})^{i} \right)=\Oh(n).\qedhere$$
\end{proof}

\subsection{Summary}\label{sec:sum}
\begin{theorem}
\label{thm:mainMC}
For any set of $b$ suffixes of $T[1..n]$, the sparse suffix tree can be
computed using $\Oh(n)$ time and $\Oh(b)$ space.
The resulting array might be incorrect with probability $n^{-c}$ for a user-defined constant $c$.
\end{theorem}
\begin{proof}
First, we apply \cref{thm:lce} to construct an efficient data structure for LCE queries.
Next, the algorithm builds the coarse compacted trie of the suffixes.
We use \cref{thm:smallb} if $b^2 \le n$ and \cref{thm:mediumb} otherwise.
Finally, we construct the sparse suffix tree using \cref{lem:uncoarse}. 
\end{proof}

\section{Las Vegas Algorithm}\label{sec:lasvegas}
A simple way to obtain a Las Vegas algorithm is to run a Monte Carlo procedure and then verify if the result is correct.
Thus, for each two adjacent suffixes $T[i..]$ and $T[j..]$ in the claimed lexicographic order,
we need to check if their longest common prefix $\ell$ has been computed correctly and
if $T[i..]$ is indeed lexicographically smaller than $T[j..]$.
Equivalently, these conditions can be stated as $T[i..i+\ell-1]=T[j..j+\ell-1]$ and
$T[i+\ell]<T[j+\ell]$.
The latter constraint is trivial to verify, so the problem boils down to checking whether $T$
satisfies a \emph{system} of $b-1$ \emph{substring equations}.

We provide a deterministic $\Oh(n\sqrt{\log b})$-time $\Oh(b)$-space algorithm for that problem,
improving upon an $\Oh(n\log b)$-time solution by I et al.\ \cite{DBLP:conf/stacs/IKK14}.
Let us start by recalling a simpler $\Oh(n\log^2 b)$-time version of their algorithm.
A straightforward reduction allows to restrict the lengths of all equations to $3\cdot 2^k$ for integer values $k$.
Now, the main idea is to relax the problem:
a YES answer is required if all equations are satisfied, but a NO---only if there is a mismatch within the first $2\cdot 2^k$ positions of an equation. 
The exact problem easily reduces to two instances of the relaxed version: 
one on the original text and one on its reverse. 
Then, the algorithm of I et al.\ works in $\Oh(\log b)$ phases. 
Each phase is given equations of length $3\cdot 2^k$ and it is responsible for making sure that there are no mismatches within the middle $2^k$ positions.
If so, the equations can be shortened to $3\cdot 2^{k-1}$. Once the maximum length is sufficiently small, all equations are verified naively.

In each phase, a graph $G$ is constructed, with nodes corresponding to blocks of the text.
Each edge represents an equation and connects nodes corresponding to blocks containing the starting positions of the involved fragments.
Then, $\Oh(|V(G)|)$ constraints are verified naively.
These are original equations forming an $\Oh(\log |V(G)|)$-spanner of the graph,
as well as constraints stating that certain fragments of the text have a given period (such constraints can also be expressed as substring equations). 
This turns out to be sufficient to implement the phase.

We introduce a few important changes to the above algorithm.
First, we avoid the eager verification of the $\Oh(|V(G)|)$  equations.  
Instead, we split them into several equations of length $3\cdot 2^{k-1}$ and process them in further phases as if they were given in the input.
Secondly, we observe that we can create these equations so that they have slack on both ends: 
just the middle $2^{k-1}$ positions need to be checked. 
Thus, we change the original relaxation to work with such equations only.  
Some technical details (e.g., the reduction from the exact problem)  become more complex, 
but now each phase simply converts a system of (relaxed) equations of length $3\cdot 2^k$ to a system of (relaxed) equations of length $3\cdot 2^{k-1}$.
This way, we achieve $\Oh(n\log b + b\log^2 n)$ running time.

The notion of difference covers lets us further exploit the slack at both sides of the equations and consequently reduce the running time to $\Oh(n\sqrt{\log b})$.
We perturb each of them so that the lengths are still uniform, but the starting positions of both fragments involved belong to a certain set.
First, this approach lets us restrict the set of starting positions of long equations, and consequently remove most of them.
(Detecting irrelevant equations also involves computing the maximum-weight spanning forest in a certain graph.)
Secondly, we reduce the number of blocks where starting positions appear, i.e., 
the number of vertices of the graph $G$ for which the spanner is constructed.
This results in fewer equations being created in each phase.

In the following sections, we formalize the above discussion.
In \cref{sec:systems} we introduce the notion of substring equation and we prove basic facts.
In \cref{sec:per,sec:span}, we formalize further tools already used by I et al., originating in combinatorics of strings and graph spanners, respectively.
Next, in \cref{sec:slow} we provide our first solution, whose
running time $\Oh(n\log b + b\log^2 n)$ is comparable to the previous state of the art, and in \cref{sec:fast} we improve the time complexity
to $\Oh(n\sqrt{\log b})$. Both algorithms require $\Oh(b)$ space.

\subsection{Systems of Substring Equations}\label{sec:systems}

Consider texts of length $n$.
For any integers $p,q,p',q'$ such that $p,p'\ge 1$, $q,q'\le n$, and  $q-p=q'-p'$, we say that $e : T[p..q]=T[p'..q']$ is a \emph{substring equation}.
The quantity $q-p+1=q'-p'+1$ is called the \emph{length} of the equation and $p'-p=q'-q$ is called the \emph{shift}.
Note that shift is \emph{oriented}: its sign is reversed when we write $e$ as $T[p'..q']=T[p..q]$.
A particular text $T$ satisfies the equation whenever $T[p..q]$ and $T[p'..q']$ are occurrences of the same string.
Equations of length 0 or less are assumed to be trivially satisfied.
For an integer $S$ we say that $T$ satisfies $e : T[p..q]=T[p'..q']$ \emph{with shortage $S$}
whenever $T$ satisfies $T[p+S..q-S]=T[p'+S..q'-S]$. 

Intuitively, our algorithm internally works with relaxed equations: 
we can verify if they are satisfied (with no shortage), but it suffices to check if they are satisfied with some shortage $S$, not necessarily uniform across all equations.
This justifies the structure of most auxiliary results.

The following fact describes how a single relaxed equation can be replaced by a system of shorter relaxed equations.
If $p=p_1$ and $q=q_1$, we can think of this procedure as splitting $e$ into $e_1,\ldots,e_m$.%
\begin{fact}\label{fct:gen_split}
Let $e, e_1,\ldots,e_m$ ($e : T[p..q]=T[p'..q'], e_i : T[p_i..q_i]=T[p'_i..q'_i]$) be substring equations with the same shift
such that $p \le p_i$ and $q_i \le q$ for $1\le i \le m$.
Moreover, let $S_i$ ($1\le i \le m$) be non-negative integers such that $p_{i+1}+S_{i+1}\le q_i-S_i+1$ for $1\le i < m$.

\begin{enumerate}[(a)]
  \item\label{it:fct:if} If $e$ is satisfied, then $e_1,\ldots,e_m$ are satisfied.
  \item\label{it:fct:of} If each $e_i$ is satisfied with shortage $S_i$, then $e$ is satisfied with shortage $S=\max(p_1-p+S_1,q-q_m+S_m)$.
\end{enumerate}
\end{fact}
\begin{proof}
(\ref{it:fct:if}) For $1\le i \le m$, $T[p_i..q_i]$ is clearly a subfragment of $T[p..q]$. Since $e_i$ has the same shift as $e$,
$T[p'_i..q'_i]$ is the corresponding subfragment of $T[p'..q']$. Thus, $e_i$ must be satisfied if $e$ is.

\noindent (\ref{it:fct:of})
We inductively prove that $T[p_1+S_1..q_1-S_1],\ldots,T[p_i+S_i..q_i-S_i]$ cover $T[p_1+S_1..q_i-S_i]$.
This claim is trivial for $i=1$, while the inductive step easily follows from the fact that $p_{i+1}+S_{i+1}\le q_i-S_i+1$.
Consequently, we obtain that $T[p_1+S_1..q_1-S_1],\ldots,T[p_m+S_m..q_m-S_m]$ cover $T[p_1+S_1..q_m-S_m]$.
By definition of $S$, the later covers $T[p+S..q-S]$. Since all equations have the same shift,
this yields (\ref{it:fct:of}).
\end{proof}

A set of substring equations on the same text $T$ is called a \emph{system}. A particular text $T$ \emph{satisfies} the system (with shortage $S$) if it satisfies all its member equations (with shortage $S$ resp.). A system is called \emph{uniform} if all equations in the system have the same length.
Uniform systems can be obtained as follows by splitting longer equations using \cref{fct:gen_split}:

\begin{corollary}\label{cor:split}
Let $e$ be an equation of length $L$. For every positive integer $\ell$, 
there exists a uniform system $E$ of $\Theta(\frac{L}{\ell})$ equations of length $\ell$ such that
if $e$ is satisfied, then $E$ is satisfied, and
if $E$ is satisfied with shortage $S$ ($S\le \frac{\ell}{3}$), then $e$ is also satisfied with shortage $S$.
Moreover, $E$ can be computed in $\Theta(\frac{L}{\ell})$ time.
\end{corollary}
\begin{proof}
Let $e : T[p..q]=T[p'..q']$ and let $r=\max(1,\floor{\frac{\ell}{3}})$.
We include in $E$ equations of length $\ell$ and start positions $(p+ir,p'+ir)$
for $0\le i < \frac{L-\ell}{r}$, as well as $(q-\ell+1,q'-\ell+1)$.
It is easy to see that these equations satisfy \cref{fct:gen_split} with $S_i = S$ provided that $2S\le \ell-r$ (e.g., if $S\le \frac{\ell}{3}$).
\end{proof}

The following result, in a simpler version introduced by I et al.~\cite{DBLP:conf/stacs/IKK14}, relates a cycle of substring equations
with periods of certain fragments.
\begin{lemma}\label{lem:shift}
Let $E=\{e_1,\ldots,e_m\}$ be a uniform system of substring equations 
with $e_i :  T[p_i..q_i]=T[p'_i..q'_i]$.
Assuming $p_{m+1}=p_1$, let us define $r = |\sum_{i=1}^m (p_{i+1}-p'_i)|$ and $R=\sum_{i=1}^m |p_{i+1}-p'_i|$.
\begin{enumerate}[(a)]
  \item\label{it:cor:if} If $E$ is satisfied, then $r$ is a period of $T[p_1+R..q_1-R]$.
  \item\label{it:cor:of} For a positive integer $S$, if $r$ is a period of $T[p_1+S..q_1-S]$
and $E\sm \{e_1\}$ is satisfied with shortage $S$,
then $e_1$ is satisfied with shortage $R+S$.
\end{enumerate}
\end{lemma}
\begin{proof}
For $i\in\{1,\ldots,m\}$ let us define $r_i = \sum_{j=1}^i (p_{j+1}-p'_j)$; we also have $r_i = \sum_{j=1}^i (q_{j+1}-q'_j)$,
since the system is uniform. Note that $r = |r_m|$ and that $|r_i|\le R$. 
Before we proceed, let us show the following auxiliary result:
\begin{claim}
If $E\sm \{e_1\}$ is satisfied with shortage $S$, then $T[p'_1+R+S..q'_1-R-S]=T[p_{i+1}-r_i+R+S..q_{i+1}-r_i-R-S]$
for each $i\in \{1,\ldots,m\}$.
\end{claim}
\begin{proof}[Proof of the claim]
We provide an inductive proof.
For $i=1$, equalities $p_2 = p'_1+r_1$ and $q_2 = q'_1+r_1$ trivially yield the inductive base.
Now, suppose that the claim is satisfied for $i-1$, i.e.,
$T[p'_1+R+S..q'_1-R-S]=T[p_{i}-r_{i-1}+R+S..q_{i}-r_{i-1}-R-S]$.
Since $|r_{i-1}|\le R$, the fact that $e_{i}$ is satisfied with shortage $S$ implies
$T[p'_1+R+S..q'_1-R-S]=T[p'_{i}-r_{i-1}+R+S..q'_{i}-r_{i-1}-R-S]$.
Equalities $p_{i+1}=p'_i + r_i$ and $q_{i+1}=q'_i + r_i$ further yield
$T[p'_1+R+S..q'_1-R-S]=T[p_{i+1}-r_{i}+R+S..q_{i+1}-r_{i}-R-S]$,
which concludes the proof of the claim.
\end{proof}
Let us proceed to the proofs of (\ref{it:cor:if}) and (\ref{it:cor:of}).

\noindent (\ref{it:cor:if})
The claim implies $T[p'_1+R..q'_1-R]=T[p_1-r_m+R..q_1-r_m-R]$,
while $e_1$ further yields $T[p_1+R..q_1-R]=T[p_1-r_m+R..q_1-r_m-R]$.
Consequently, $r=|r_m|$ must be a period of $T[p_1+R..q_1-R]$.

\noindent (\ref{it:cor:of})
The claim implies
$T[p'_1+R+S..q'_1-R-S]=T[p_1-r_m+R+S..q_1-r_m-R-S]$.
The fact that $T[p_1+S..q_1-S]$ has period $r=|r_m|\le R$, yields
$T[p_1-r_m+R+S..q_1-r_m-R-S]=T[p_1+R+S..q_1-R-S]$.
Combining these two equalities, we conclude that
$T[p'_1+R+S..q'_1-R-S] = T[p_1+R+S..q_1-R-S]$,
i.e., that $e_1$ is satisfied with shortage $R+S$.
\end{proof}

\subsection{Period constraints}\label{sec:per}
The fact that a fragment $T[p..q]$ has period $r$ can be expressed as a substring equation
$T[p..q-r]=T[p+r..q]$. 
This condition is particularly important if $2r \le |T[p..q]|$,
i.e., if $r$ does not exceed the length of the underlying equation.
Thus, if for an equation $e: T[p..q]=T[p'..q']$ the absolute value of the shift does not exceed the length, we say that the equation is a \emph{(substring) period constraint}
 which \emph{concerns} $T[\min(p,p')..\max(q,q')]$ and \emph{enforces} period $|p-p'|$.
Periodicity Lemma, a classic tool of combinatorics of words, 
lets us essentially reduce several period constraints enforcing different periods with 
a single period constraint enforcing the greatest common divisor of these periods.

\begin{lemma}[Periodicity Lemma, Fine and Wilf~\cite{PeriodicityLemma}]\label{lem:perlem}
If a string $x$ has length $n$ and periods $p,q$ such that $n\ge p+q-\gcd(p,q)$,
then it also has $\gcd(p,q)$ as a period.
\end{lemma}

\begin{lemma}\label{lem:per}
Let $E=\{e_1,\ldots,e_m\}$ be a system of period constraints concerning a fragment $T[p..q]$ and enforcing periods $r_1,\ldots,r_m$, 
and let $e$ be a constraint enforcing $r=\gcd(r_1,\ldots,r_m)$ as a period of $T[p..q]$.
\begin{enumerate}[(a)]
  \item\label{it:per:if} $E$ is satisfied if and only if $e$ is satisfied.
  \item\label{it:per:of} For every positive integer $S$, if $e$ is satisfied with shortage $S$, then $E$ is satisfied with shortage $S$.
\end{enumerate}
\end{lemma}
\begin{proof}
Note that $e: T[p..q-r]=T[p+r..q]$ is satisfied with shortage $S$ whenever $T[p+S..q-S]$ has period $r$ (not necessarily proper).
Thus, the fact that $e$ is satisfied with shortage $S$ means that $T[p+S..q-S]$ has period $r$ and consequently all periods $r_i$ (as $r\mid r_i$).
Consequently, $E$ is satisfied with storage $S$, as claimed in (\ref{it:per:of}).
The `if' part of (\ref{it:per:if}) is a special case for $S=0$, so let us proceed to the proof of the `only if' part.
Suppose that $E$ is satisfied. 
Iteratively applying Periodicity Lemma (\cref{lem:perlem}) to $w=T[p..q]$ with periods $r_i$ and $\gcd(r_1,\ldots,r_{i-1})$, we conclude
that $\gcd(r_1,\ldots,r_m)$ is a period of $T[p..q]$, i.e., that $e$ is satisfied. 
Note that $\gcd(r_1,\ldots,r_{i-1})+r_i \le r_{i-1}+r_i \le |T[p..q]|$ due to the fact
that $r_i \le \frac{1}{2}|T[p..q]|$ by definition of a period constraint.
\end{proof}

\subsection{Graph Spanners}\label{sec:span}
In this section, we recall a classic construction of a $(2\floor{\log n}-1)$-multiplicative graph spanner of size at most $2n$.
The underlying original idea is by Awerbuch~\cite{DBLP:journals/jacm/Awerbuch85}.
These spanners have already been implicitly applied in \cite{DBLP:conf/stacs/IKK14} to verify substring equations.
Our setting requires some extra information concerning the paths witnessing that each edge has low enough stretch.
Thus, we formulate the algorithm in a non-standard way, formalizing the contribution of \cite{DBLP:conf/stacs/IKK14}.

Let $G=(V,E)$ be a undirected multigraph. We interpret each edge of $G$ as a pair of inverse arcs.
For a subset $F\sub E$, we define $\vec{F}$ to be the set of arcs corresponding to the edges in $F$.
For an arc $e\in \vec{E}$, we define $e^R$ as the arc inverse to $e$ (this way $(e^R)^R = e$).
We say that $w : \vec{E}\to \mathbb{Z}$ in an \emph{oriented weight function} if $w(e^R)=-w(e)$ for each $e\in \vec{E}$.
The oriented weight function can be naturally extended to (oriented) paths and cycles: we set $w(C)=\sum_{e\in C} w(e)$.

\begin{lemma}[\cite{DBLP:conf/stacs/IKK14}]\label{lem:graph}
Let $G=(V,E)$ be an undirected multigraph with oriented weight function $w$. 
In $\Oh(|V|+|E|)$ time, one can compute a subset $E'\sub E$ of size at most $2|V|$
and an oriented weight function $c$ which satisfy the following condition:
for each arc $e\in \vec{E}$, there exists a directed cycle $C \sub \vec{E'}\cup \{e\}$ containing $e$ such that 
  $|C|\le 2\log |V|$ and $c(e)=w(C)$.
\end{lemma}
\begin{proof}
Let $v$ be an arbitrary vertex and, for a non-negative integer $i$, let $V_{i}$ be the set of vertices
whose (unweighted) distance from $v$ is at most $i$.
Let $d$ be the smallest positive such that $|V_{d}|\le 2|V_{d-1}|+1$.
Note that $|V_{i}| \ge 2|V_{i-1}|+2$ for $i < d$ and thus $d \le \floor{\log |V|}$.

Observe that breadth-first search starting from $v$ can be used to determine $d$ and $V_{d}$.
Moreover, it results in a spanning tree of $G[V_{d}]$ rooted at $v$; we shall include the tree-edges in $E'$.
Each arc $e$ incident to at least one vertex in $V_{d-1}$ can be completed to a cycle $C$ using a tree-path of length at most $2d-1$.
Note that after linear-time preprocessing of the tree (to compute the weight $w(P_u)$ of the tree-path from $u$ to $v$ for every vertex $u\in V_d$),
the weight $w(C)$ can be computed in constant time (as $w(e)+w(P_{u'})-w(P_u)$ for an arc $e$ from $u$ to $u'$). We shall set $c(e)=w(C)$ for these arcs.

The procedure above (including BFS) runs in time proportional to the number of arcs incident to $V_{d-1}$,
and it lets us remove $V_{d-1}$ from the graph. The number of edges added to $E'$ is at most $|V_{d}|-1\le 2|V_{d-1}|$,
i.e., twice as much as the number of removed vertices.
Thus, if we recursively continue the same procedure, we get a set $E'$ of size at most $2|V|$ and an oriented weight function $c$
satisfying the desired condition.
\end{proof}

\subsection{$\Oh(n\log b + b\log^2n)$-time Solution}\label{sec:slow}

For a uniform system $E$ of equations of length $L$ and \emph{block size} $B$,
we construct a multigraph graph $G_{E,B}$ with $\Oh(\frac{n}{B})$ vertices and edge set identified with $E$,
and an oriented weight $w$ function corresponding to the shift of the equations.

Vertices of $G_{E,B}$ correspond to consecutive blocks of $T$ starting at \emph{aligned} positions of the form $1+kB$,
and are represented by these positions.
For a position $p$, we denote the preceding aligned position by $\pred(p)$ ($p - B < \pred(p) \le p$).
An equation $e : T[p..q]=T[p'..q']$ is represented by an edge between $\pred(p)$ and $\pred(p')$.
The weights of the underlying arcs are $p'-p$ (for the arc from $\pred(p)$) and $p-p'$ (for the arc from $\pred(p')$).
We discard isolated vertices from $G_{E,B}$, which lets us store this graph in $\Oh(|E|)$ space.

\begin{observation}\label{obs:graph_constr}
The graph $G_{E,B}$ and the weight function $w$ can be constructed in $\Oh(|E|+\frac{n}{B})$ time.
\end{observation}

A spanner of the graph $G_{E,B}$, combined with \cref{lem:shift,lem:per}, can be used to reduce a uniform system of relaxed substring equations
to a system of shorter relaxed substring equations. This result generalizes the main concept of the $\Oh(n\log^2b)$-time algorithm of I et al.~\cite{DBLP:conf/stacs/IKK14}.
\begin{lemma}\label{lem:eq_graph}
Let $E$ be a system of substring equations of length $L$ on a text of length $n$.
For a fixed block-size $B$,  let $N$ be the number of vertices of the graph $G_{E,B}$.
Moreover, let $S$ be a positive integer such that $B(2\floor{\log N} + 1) \le S \le \frac{L}{4}$.
There is a system $\bar{E}$ of $\Oh(N)$ equations of length between $L-3S$ and $L$ such that:
\begin{enumerate}[(a)]
  \item if $E$ is satisfied, then $\bar{E}$ is satisfied,
  \item if $\bar{E}$ is satisfied with shortage $S'$, then $E$ is satisfied with shortage $S'+2S$.
\end{enumerate}
Moreover, such $\bar{E}$ can be computed in $\Oh(|E|\log S)$ time and $\Oh(|E|)$ space
given the graph $G_{E,B}$.
\end{lemma}
\begin{proof}
For every vertex (represented by aligned position $p$), we define a \emph{canonical} fragment $T_p = T[p+S..p+L-1-S]$.
We apply \cref{lem:graph} to $G_{E,B}$, which results in a subset $E'\sub E$ of size $2N$ and an oriented weight function $c$.
For every (oriented) equation $e:T[p..q]=T[p'..q']$ in $E$, we consider a period constraint $e''$ enforcing $|c(e)|$ as a period of $T_{\pred(p)}$.

\begin{claim}
If $E$ is satisfied, then $e''$ is also satisfied. 
Moreover, if $E'\cup \{e''\}$ is satisfied with shortage $S'$,
then $e$ is satisfied with shortage $S'+2S$.
\end{claim}
\begin{proof}
Let $C=(e_1,\ldots,e_m)$ be the witness cycle for $e$ with $e=e_1$ and $e_i : T[p_i..q_i]=T[p'_i..q'_i]$.
We shall apply \cref{lem:shift} for equations in $C$. 
Note that $c(e)=\sum_{i=1}^m (p'_i - p_i)=\sum_{i=1}^m (p'_i-p_{i+1})$ and thus $|c(e)|=r$. 
Moreover, since $\pred(p_{i+1})=\pred(p'_i)$, we have $|p'_i-p_{i+1}|\le B$,
i.e., $r \le R \le |C|B$. Since $|C|\le 2\floor{\log N}$, this yields $S\ge R+B$.

Let $T_{\pred(p)}=T[\bar{p}..\bar{q}]$. 
We shall first prove that $T_{\pred(p)}$ is contained in $T[p+R..q-R]$ and contains $T[p+2S-R..q-2S+R]$.
First, note that $\bar{p}=\pred(p)+S\le p+S \le p+2S-R$ 
and $\bar{q}=\pred(p)+L-1-S \le p+L-1-S=q-S \le q-R$ since $S\ge R$.
Similarly, $\bar{p}=\pred(p)+S \ge p-B+S \ge p+R$
and $\bar{q}=\pred(p)+L-1-S \ge p+L-1-B-S = q-S-B\ge q-2S+R$
since $S\ge B+R$.

Suppose that $E$ is satisfied. Since $C\sub E$,
\cref{lem:shift}(\ref{it:cor:if}) implies that $T[p+R..q-R]$, and thus also $T_{\pred(p)}$, has period $r$.
Hence, $e''$ is satisfied.

Now, suppose that $E'\cup\{e''\}$ is satisfied with shortage $S'$.
This yields that $C\sm \{e\}$ is satisfied with shortage $S'+2S-R$ and that $T[\bar{p}+S'..\bar{q}-S']$,
i.e., also $T[p+(S'+2S-R)..q-(S'+2S-R)]$, has period $r$.
Consequently, \cref{lem:shift}(\ref{it:cor:of}) implies that $e$ is satisfied with shortage $R+(S'+2S-R)=S'+2S$.
\end{proof}
Finally, note that the shift of the equation $e''$ is $|c(e)|\le S$, while the length is $L-2S-c(e)\le L-3S$.
Since we assumed that $S \le \frac{L}{4}$, the equation $e''$ is indeed a period constraint concerning a canonical fragment.

Hence, the system $E'\cup E''$ with $E''=\{e'' : e\in E\}$ satisfies the claimed conditions on $\bar{E}$
except that that its size is $|E|+\Oh(g)$ rather than $\Oh(g)$. 
For each canonical fragment, \cref{lem:per} lets us replace all period constraints concerning this fragment with a single constraint. 
As far as the implementation is concerned, this requires a single $\gcd$ query per period constraint.
Since the periods are bounded by $S$, Euclid's algorithm takes $\Oh(\log S)$ time, which gives
$\Oh(|E|\log S)$ total running time.
\end{proof}

\begin{corollary}\label{cor:reduce}
Let $k\ge 2$ be a positive integer, and let $E$ be a system of substring equations of length $3\cdot 2^k$ on a text $T$ of length $n$.
There exists a system $E'$ of $\Oh(\frac{n}{2^k}\log \frac{n}{2^k})$ substring equations of length $3\cdot 2^{k-1}$ such that:
\begin{enumerate}[(a)]
  \item if $E$ is satisfied, then $E'$ is satisfied;
  \item if $E'$ is satisfied with shortage $2^{k-1}$, then $E$ is satisfied with shortage $2^k$.
\end{enumerate}
Moreover, such $E'$ can be computed in $\Oh(|E|k+\frac{n}{2^k}\log \frac{n}{2^k})$ time and $\Oh(|E|+\frac{n}{2^k}\log \frac{n}{2^k})$  space.
\end{corollary}
\begin{proof}
We set $L = 3\cdot 2^k$, $S=2^{k-2}$, and the block size $B$ as the largest integer such that $B(2\ceil{\log \frac{n}{B}}+1) \le S$.
We construct $G_{E,B}$ using \cref{obs:graph_constr}.
The number of vertices satisfies $N = \ceil{\frac{n}{B}} = \Theta(\frac{n}{2^k}\log \frac{n}{2^k})$.
We have $B(2\log \ceil{N}+1)\le S \le \frac{L}{4}$, so we can apply \cref{lem:eq_graph}
to obtain a set system $\bar{E}$ containing $\Oh(N)$ equations of length between $L-3S=9\cdot 2^{k-2}$ and $L=12\cdot 2^{k-2}$.

Next, we use \cref{cor:split} for each equation in $\bar{E}$ (with $\ell=3\cdot 2^{k-1}$) to obtain a system $E'$
consisting of equations of length $3\cdot 2^{k-1}$. 
Clearly, if $E$ is satisfied, then also $\bar{E}$ and $E'$ are. On the other hand, if $E'$ is satisfied with shortage $2^{k-1}$,
then $\bar{E}$ is satisfied with shortage $2^{k-1}$, and consequently $E$ is satisfied with shortage $2^{k-1}+2\cdot 2^{k-2}=2^k$,
as claimed. 

The running time $\Oh(|E|\log S+N)$ and the space consumption $\Oh(|E|+N)$ are as claimed.
\end{proof}

The following result, complementary to \cref{cor:reduce}, when applied recursively, lets us reduce a given system of substring equations
to $\Oh(\log n)$ systems of uniform relaxed systems.

\begin{lemma}\label{lem:decompose}
Let $k$ be a non-negative integer and let $E$ be a system of $b$ substring equations on a text $T$, each of length at least $3\cdot 2^k$.
In $\Oh(b)$ time one can output systems $E'$ and $E''$ such that
\begin{enumerate}[(a)]
  \item $E'$ contains $\Oh(b)$ equations, each of length exactly $3\cdot 2^k$;
  \item $E''$ contains at most $b$ equations, each of length at least $3\cdot 2^{k+1}$;
  \item\label{it:three} if $E$ is satisfied, then $E'$ are $E''$ are also satisfied;
  \item\label{it:four} if $E'$ is satisfied with shortage $S$, $S\le 2^k$, and $E''$ is satisfied with shortage $2^{k+1}$, then $E$ is satisfied with shortage $S$.
\end{enumerate}
\end{lemma}
\begin{proof}
Note that it suffices to prove the lemma for $b=1$; for $b>1$, we construct $E'$ and $E''$
as unions of the systems generated for each equation $e\in E$. Thus, we assume that $E$ consists of a single equation $e$.

Let $L$ be the length of $e$. If $L < 3\cdot 2^{k+1}$, then we apply \cref{cor:split}, return the resulting system as $E'$,
and set $E''=\emptyset$. Conditions (\ref{it:three}) and (\ref{it:four}) are satisfied because $S\le 2^k=\frac{1}{3}\cdot 3\cdot 2^k$.

Thus, we may assume that $e: T[p..q]=T[p'..q']$ has length $L\ge 3\cdot 2^{k+1}$.
We consider a sequence of three equations $e_1 :  T[p..p+3\cdot 2^k-1]=T[p'..p'+3\cdot 2^k -1] , e_2=e, e_3=T[q-3\cdot 2^k+1..q]=T[q'-3\cdot 2^k+1..q']$
with shortages $S_1=S$, $S_2=2^{k+1}$, and $S_3=S$.
Since $S\le 2^k$, the sequence clearly satisfies \cref{fct:gen_split}, which yields  (\ref{it:three}) and (\ref{it:four}).
\end{proof}

Now, we are ready to describe the first version of our verification algorithm.

\begin{theorem}\label{thm:slowLV}
A system $E$ of $b$ substring equations on a text $T$ of length $n$ can
be verified in $\Oh(n\log b + b\log^2 n)$ time using $\Oh(b)$ space.
\end{theorem}
\begin{proof}
We set $\ell = \floor{\log \frac{n\log b}{b}}$ and naively check all equations shorter than $3\cdot 2^{\ell}$. 
This takes $\Oh(b\cdot \frac{n\log b}{b})=\Oh(n\log b)$ time. From now on, we assume that all equations are of length at
least $3\cdot 2^{\ell}$.

The equations are processed iteratively applying \cref{lem:decompose} for increasing values of $k$, starting from $k=\ell$,
which results in systems $E_{\ell},\ldots,E_r$ ($r=\floor{\log\frac{n}{3}}$), such that each system $E_k$,
$\ell \le k \le r$, contains $\Oh(b)$ equations, each of length $3\cdot 2^k$.
Moreover, condition (\ref{it:three}) of \cref{lem:decompose} implies that if the system $E$ is satisfied, then all systems $E_k$
is satisfied. Moreover, by condition (\ref{it:four}), if $E_\ell$ is satisfied and each $E_k$, $\ell < k \le r$,
is satisfied with shortage $2^{k}$, then the input system $E$ is satisfied.
Note that the systems $E_k$ together do not fit in $\Oh(b)$ space, but each of them can be generated in $\Oh(b(r-\ell+1))=\Oh(b\log n)$
time.

We process systems $E_k$ for decreasing values $k$. We define $F_{r}=\emptyset$ and for $\ell\le k < r$,
we define $F_{k}$ as the effect of application of the reduction of \cref{cor:reduce} to $E_{k+1}\cup F_{k+1}$.
Note that $F_k$ is a uniform system with $\Oh(b + \frac{n}{2^k}\log \frac{n}{2^k})$ equations of length $3\cdot 2^k$.
Since $\frac{n}{2^k}=\Theta(\frac{b}{\log b})$, we actually have $|F_k|=\Oh(b)$ for each $k$.

We shall prove that $E$ is satisfied if and only if $E_{\ell}$ and $F_{\ell}$ are both satisfied.
The `only if' part is easier: we have already observed that that systems $E_k$ are satisfied for $\ell \le k \le r$.
The fact that the systems $F_k$ are satisfied is proved by induction for decreasing $k$. For $k=r$ this is trivial.
For $k<r$, we have that both $E_{k+1}$ and $F_{k+1}$ are satisfied, and \cref{cor:reduce} implies that $F_k$ must also be satisfied.
In the `if' part we inductively prove that the systems $E_k$ and $F_k$ are satisfied with shortage $2^k$, this time
for increasing values of $k$. For $k=\ell$ this is trivial, since they are actually satisfied with no shortage.
For $k>\ell$, we deduce from \cref{cor:reduce} that the fact that $F_{k-1}$ is satisfied with shortage $2^{k-1}$
yields that $F_{k}\cup E_{k}$ is satisfied with shortage $2^{k}$.
Finally, by the analysis in the first part of the proof, we note that $E$ must be satisfied, as claimed.

Consequently, in order to verify if $E$ is satisfied, we can equivalently verify $E_\ell\cup F_\ell$.
The latter systems contains $\Oh(b)$ equations, each of length $\Oh(\frac{n\log b}{b})$, so this takes $\Oh(n\log b)$ time and $\Oh(b)$ space.
The construction of each $F_k$ from $F_{k+1}$ takes $\Oh(b)$ space and  $\Oh(b\log n)$ time, which sums up to $\Oh(b\log^2 n)$ in total.
\end{proof}

\subsection{$\Oh(n\sqrt{\log b})$-time solution}\label{sec:fast}
We reduce the running time of the algorithm by applying difference covers to improve \cref{cor:reduce} and \cref{lem:decompose}.
We start with a counterpart of \cref{cor:reduce}.

\begin{lemma}\label{lem:reduce2}
Let $k\ge 3$ be a positive integer, and let $E$ be a system of substring equations of length $3\cdot 2^k$ on a text $T$ of length $n$.
There exists a system $E'$ of $\Oh(\frac{n}{2^k}\sqrt{\log \frac{n}{2^k}})$ substring equations of length $3\cdot 2^{k-1}$ such that:
\begin{enumerate}[(a)]
  \item if $E$ is satisfied, then $E'$ is satisfied;
  \item if $E'$ is satisfied with shortage $2^{k-1}$, then $E$ is satisfied with shortage $2^k$.
\end{enumerate}
Moreover, such $E'$ can be computed in $\Oh(|E|k+\frac{n}{2^k}\sqrt{\log \frac{n}{2^k}})$ time and $\Oh(|E|+\frac{n}{2^k}\sqrt{\log \frac{n}{2^k}})$  space.
\end{lemma}
\begin{proof}
We set $L=3\cdot 2^k$, $S=2^{k-3}$, and the block size $B$ as the largest integer such that $B(2\ceil{\log \frac{n}{B}}+1)\le S$.
However, before building the graph $G_{E,B}$, we construct a $\floor{\frac{S}{B}}$-difference cover of $\{1,\ldots,\ceil{\frac{n}{B}}\}$
using \cref{lem:cover}; its size is $\Oh(\frac{n}{\sqrt{SB}})$. We slightly abuse the notation and define the shift function $h$ so that its arguments are aligned positions, and the values are multiples of $B$.

We transform $E$ into a system $\tilde{E}$ of equations of length $L-S$:
for an equation $e : T[p..q]=T[p'..q']$  we compute $h_e = h(\pred(p),\pred(p'))$
and construct $e' : T[p+h_e..q+h_e-S] = T[p'+h_e .. q'+h_e-S]$. Clearly, if $e$ is satisfied, then $e'$ is satisfied,
and if $e'$ is satisfied with shortage $2^{k-1}$, then $e$ is satisfied with shortage $2^{k-1}+2^{k-3}$.

After this operation, we are guaranteed that there are only $N=\Oh(\sqrt{\frac{n}{S}}\cdot \sqrt{\frac{n}{B}})=\Oh(\frac{n}{S}\sqrt{\log\frac{n}{S}})$ (non-isolated) vertices in $G_{\tilde{E},B}$.
Moreover, the fact that the difference cover can be indexed efficiently, implies that $G_{\tilde{E},B}$ can be constructed in $\Oh(|E|+\frac{n}{S}\sqrt{\log\frac{n}{S}})$time. 
Note that $B(2\ceil{\log\frac{n}{B}}+1)\le B(2\ceil{\log N}+1)\le S \le \frac{L-S}{4}$, so we may apply \cref{lem:eq_graph} to $\tilde{E}$,
which results in a system $\bar{E}$.

The lengths of equations in $\bar{E}$ are between $L-4S=20\cdot 2^{k-3}$ and $L-S=23\cdot 2^{k-3}$;
we apply \cref{cor:split} to each of them in order to obtain a uniform system $E'$ with equations of length $3\cdot 2^{k-1}$.

If $E$ is satisfied, then $\tilde{E}$, $\bar{E}$, and $E'$ are all satisfied. 
On the other hand, if $E'$ is satisfied with shortage $2^{k-1}$, then $\bar{E}$ is satisfied with shortage $2^{k-1}$,
$\tilde{E}$ is satisfied with shortage $2^{k-1}+2\cdot 2^{k-3}$, and $E$ is satisfied with shortage $2^{k-1}+3\cdot 2^{k-3}\le2^k$,
as claimed.
\end{proof}

Before we state a stronger version of \cref{lem:decompose}, let us show an auxiliary result, based on the notion of the minimum spanning forest of a graph.
\begin{lemma}\label{lem:mst}
Let $E$ be a system of $b$ substring equations on a text $T$ of length $n$ and let $DC$ be a
subset of $\{1,\ldots,n\}$ which can be indexed efficiently and contains $p$ and $p'$ for each $e : T[p..q]=T[p'..q']$ in $E$.
In $\Oh(b+|DC|)$ time we can compute a subsystem $E'\sub E$ of size at most $|DC|$, such that for any shortage $S$, 
$E'$ is satisfied with shortage $S$ if and only if $E$ is satisfied with shortage $S$.
\end{lemma} 
\begin{proof}
We construct a weighted undirected multigraph $G$ with vertex set $DC$.
Each equation $e : T[p..q]=T[p'..q']$ is represented as an edge between $p$ and $p'$ with weight $q-p+1=q'-p'+1$
equal to the length of the equation. The graph can be build in $\Oh(b+|DC|)$ time, since $DC$ can be indexed efficiently.

We compute a maximum-weight spanning forest of $G$ (using a linear-time algorithm of Fredman and Willard~\cite{DBLP:journals/jcss/FredmanW94})
and return the subsystem $E'$ corresponding to the edges of the forest.

It suffices to prove that equations in $E\sm E'$ can be dropped. For such an equation $e:T[p..q]=T[p'..q']$ of length $L$,
there is a sequence of equations $e_1,\ldots,e_m\in E'$ ($e_i  : T[p_i..q_i]=T[p'_i..q'_i]$) of length at least $L$
such that $p_1 = p$, $p'_m = p'$, and $p_{i+1}=p'_i$ for $1\le i < m$.
Clearly, if $E'$ is satisfied with shortage $S$, then $e_1,\ldots,e_m$ are all satisfied with shortage $S$,
and consequently $T[p_i+S..p_i+L-1-S]=T[p'_i+S..p'_i+L-1-S]=T[p+S..q-S]$ for each $i\in \{1,\ldots,m\}$;
in particular $T[p'+S..q'-S]=T[p+S..q-S]$, i.e., $e$ is satisfied with shortage $S$.
\end{proof}

\begin{lemma}\label{lem:decompose2}
Let $k$ be a non-negative integer and let $E$ be a system of $b$ substring equations on a text $T$, each of length at least $3\cdot 2^k$.
In $\Oh(b)$ time one can output systems $E'$ and $E''$ such that
\begin{enumerate}[(a)]
  \item $E'$ contains $\Oh(b)$ equations, each of length exactly $3\cdot 2^k$;
  \item $E''$ contains $\min(b,\Oh(\frac{n}{\sqrt{2^{k+1}}}))$ equations, each of length at least $3\cdot 2^{k+1}$;
  \item\label{it:three2} if $E$ is satisfied, then $E'$ are $E''$ are also satisfied;
  \item\label{it:four2} if $E'$ is satisfied with shortage $S$, $S\le 2^k$, and $E''$ is satisfied with shortage $2^{k+1}$, then $E$ is satisfied with shortage $S$.
\end{enumerate}
\end{lemma}
\begin{proof}
If $b \le \frac{n}{\sqrt{2^{k+1}}}$, we simply fall back to \cref{lem:decompose}.
Otherwise, like in the proof of \cref{lem:decompose}, we use a different approach for and short long equations $e\in E$.
This time, however, we set a larger threshold of $4\cdot 2^{k+1}$ to distinguish between the two cases.
If the length $L$ of the equation $e\in E$ is below this value,
we apply \cref{cor:split} and insert the resulting equations of length $3\cdot 2^k$ to $E'$.

To handle longer equations $e : T[p..q]=T[p'..q']$, we use a $2^{k+1}$-difference-cover $DC$ of $\{1,\ldots,n\}$ constructed according to \cref{lem:cover}.
We generate a sequence of substring equations $e_1,\ldots,e_5$.
The equation $e_4 : T[p+h(p,p')..q]=T[p'+h(p,p')..q]$ is inserted to $E''$, and the remaining equations, inserted to $E'$, are defined to have length $3\cdot 2^k$ and start positions $(p,p')$, $(p+2^k,p'+2^k)$, $(p+2^{k+1},p'+2^{k+1})$, and $(q-3\cdot 2^{k}+1, q'-3\cdot 2^{k}+1)$, respectively.
Since $p+h(p,p')\le p+2^{k+1}\le 1+p+2^{k+1}+ (3\cdot 2^{k}-1)-S-2^{k+1}$,
we can use \cref{fct:gen_split} to prove that (\ref{it:three2}) and (\ref{it:four2}) are satisfied.

Finally, we reduce $E''$ using \cref{lem:mst}. Since the starting positions belong to $DC$, this results
in $\Oh(\frac{n}{\sqrt{2^{k+1}}})$ equations. The running time of the reduction is $\Oh(b)$, as claimed.
\end{proof}

Having developed all necessary tools, we are ready to prove the main results of this section.

\begin{theorem}\label{thm:LV}
A system $E$ of $b$ substring equations on a text $T$ of length $n$ can
be verified in $\Oh(n\sqrt{\log b})$ time using $\Oh(b)$ space.
\end{theorem}
\begin{proof}
We set $\ell = \floor{\log \frac{n\sqrt{\log b}}{b}}$ and naively check all equations shorter then $3\cdot 2^{\ell}$. 
This takes $\Oh(b\cdot \frac{n\sqrt{\log b}}{b})=\Oh(n\sqrt{\log b})$ time. 
From now on, we assume that all equations are of length at least $3\cdot 2^{\ell}$.

The equations are processed iteratively applying \cref{lem:decompose2} for increasing values of $k$, starting from $k=\ell$.
This results in systems $E_{\ell},\ldots,E_r$ ($r=\floor{\log\frac{n}{3}}$), such that each system $E_k$,
$\ell \le k \le r$, contains $\Oh(\min(b,\frac{n}{\sqrt{2^k}}))$ equations, each of length $3\cdot 2^k$.
Moreover, condition (\ref{it:three}) of \cref{lem:decompose2} implies that if the system $E$ is satisfied, then all systems $E_k$
is satisfied. By condition (\ref{it:four}), if $E_\ell$ is satisfied and each $E_k$, $\ell < k \le r$,
is satisfied with shortage $2^{k}$, then the input system $E$ is satisfied.

Let $m=2\floor{\log\frac{n}{b}}$.
Note that the systems $E_k,\ldots,E_m$ together do not fit in $\Oh(b)$ space, but each of them can be generated in $\Oh(b(m-\ell+1))=\Oh(b\log \frac{n}{b})$
time. On the other hand, the bound $\Oh(\frac{n}{\sqrt{2^k}})$ on the size of $E_{m+1},\ldots,E_r$ decreases geometrically an thus their total size is 
$\Oh(\frac{n}{\sqrt{2^m}})=\Oh(b)$. The time required to generate these systems is also $\Oh(b)$.

We process systems $E_k$ for decreasing values $k$. We define $F_{r}=\emptyset$ and for $\ell\le k < r$,
we define $F_{k}$ as the effective application of the reduction of \cref{lem:reduce2} to $E_{k+1}\cup F_{k+1}$.
Note that $F_k$ is a uniform system with $\Oh(\min(b,\frac{n}{\sqrt{2^k}}) + \frac{n}{2^k}\sqrt{\log \frac{n}{2^k}})$ equations of length $3\cdot 2^k$.
Since $\frac{n}{2^k}=\Theta(\frac{b}{\sqrt{\log b}})$, we actually have $|F_k|=\Oh(\min(b,\frac{n}{\sqrt{2^k}}))$ for each $k$.

The proof that $E$ is satisfied if and only if $E_{\ell}$ and $F_{\ell}$ are both satisfied
is the same as in \cref{thm:slowLV}; we just use \cref{lem:reduce2} instead of \cref{cor:reduce}.
The system $F_\ell\cup E_\ell$ contains $\Oh(b)$ equations, each of length $\Oh(\frac{n\sqrt{\log b}}{b})$, so its verification takes $\Oh(n\sqrt{\log b})$ time and $\Oh(b)$ space.
The construction of each $F_k$ from $F_{k+1}$ takes $\Oh(b)$ space and $\Oh(b\log\frac{n}{b})$ time for $k\le m$,
and $\Oh(b\log \frac{n}{b})$ time in total for $k\ge m$.
This gives overall running time of $\Oh(b\log^2 \frac{n}{b})=\Oh(n)$.
\end{proof}
\begin{corollary}
\label{cor:mainLV}
The sparse suffix tree of any $b$ suffixes of a text of length $n$
can be computed using $\Oh(n\sqrt{\log b})$ time and $\Oh(b)$ space.
The algorithm returns $\bot$ w.\ prob.\ $n^{-c}$ for user-defined constant~$c$.
\end{corollary}

\bibliographystyle{plainurl}
\bibliography{sparse}

\appendix
\newpage

\end{document}